\documentstyle[multicol,aps,epsf]{revtex}
\begin{document}

\sloppy

\title{Properties of the strongly correlated two-dimensional electron gas
in Si MOSFETs.}

\author{B.Spivak}

\address{Physics Department, University of Washington, Seattle, WA 98195}

\maketitle

\begin{abstract} We discuss properties of the strongly correlated
two-dimensional electron gas in Si MOSFETs at low concentrations assuming
that the
electron liquid is close to crystallization. 
Analogy with the theory of ${}^3{\rm He}$ is emphasized. 

\end{abstract}


\begin{multicols}{2}

\section{Introduction}

Recent
experiments~
\cite{krav,sar,kravHpar,kravHtilt,pudHpar,shah,pudcam,vit,klap1,klap2}
in 
high-mobility 
two-dimensional electron systems indicate that 
in the presence of disorder they cannot be described 
by the conventional single particle localization theory~\cite{4and}. 
A number of unusual features have been reported taking place in the 
dependence of
kinetic coefficients
on the magnetic field $H$ and temperature $T$. They are observed
 at low $T$ and relatively small
electron concentration $n$.
This regime corresponds to $r_{s}=E_{p}/E_{F}\gg 1$, where
$E_{p}=e^{2}p_{F}/\epsilon\hbar$ is the potential energy
per one electron,  $E_{F}=p_{F}^{2}/2m$ is the Fermi energy,
$\epsilon$ is the dielectric constant, $m$ is the bare electron mass 
and $p_{F}$ is the Fermi momentum.
In this article we give a qualitative interpretation of 
the experimental results
assuming that the electron liquid is close to Wigner 
crystallization. Our analysis is based on an analogy with 
the theory of ${}^3{\rm He}$.
We would like to explain the following experimentally observed facts.

A. The electron system exhibits a ``transition" as a 
function of $n$ from a 
metallic phase, where the resistivity $\rho$ of the system
saturates at low temperatures, to an insulating phase, where the
resistivity increases as $T$ 
decreases. 
The value of the critical concentration $n_{c}$ depends on the amount of
disorder in the sample and corresponds to $r_{s}=r^{c}_{s}\gg 1$.
On the metallic side of the transition at $T=0$ and $n\approx n_{c}$ the
value of the resistance is of order $\hbar/e^{2}$.

B. At $T=0$ and electron concentration sufficiently close 
to critical, magnetic field parallel to the film $H_{\|}$ suppresses the metallic
phase and
drives the system toward the insulating
phase~\cite{kravHpar,pudHpar,pudanis}. 
Thus the critical metal-insulator concentration $n_{c}(H_{\|})$ increases with
$H_{\|}$.

 In the metallic phase ($n>n_{c}(H_{\|}=0)$) and at small $T$ the system 
exhibits a big
positive magnetoresistance as a function of $H_{\|}$.   
This magnetoresistance
saturates at 
$H_{\|}\ge H_{\|}^{c}(n)$ and $\rho(H_{\|}^{c})/\rho(0)\gg 
1$~\cite{kravHpar,pudHpar}.

C. In the dielectric phase the system exhibits a big positive magnetoresistance
as a
function of $H_{\|}$, which saturates at $H_{\|}>H_{\|}^{c}$.
However, at given $H_{\|}>H_{\|}^{c}$ the conductance of
the system exhibits a big negative 
magnetoresistance as a function of the component of the magnetic field 
$H_{\bot}$ perpendicular to the film~\cite{kravHtilt}.

D. In the metallic phase at $H_{\|}=0$ and $T<E_{F}$ the  resistance
$\rho(T)$ significantly
increases with temperature.  The characteristic value of
$d \ln\rho/d T>E_{F}^{-1}$ at small $T$ is large 
and depends on the value of $n-n_{c}$.

E. If at $H_{\|}>H_{\|}^{c}$ the system is still in the metallic phase 
($n>n_{c}(H_{\|})$), the $T$ dependence of the resistance is much smaller
than in the $H_{\|}=0$ case
$(d \ln\rho(H_{\|}>H_{\|}^{c})/d T\ll E_{F}^{-1})$~\cite{vit,klap1}.

F. In the metallic regime in magnetic fields of order
$H^{c}_{\|}$ or less the system exhibits a large anisotropy: The
magnetoresistance in the parallel field is 
much larger than the magnetoresistance in the perpendicular
field~\cite{kravHtilt}.

\section{A comparison of low temperature properties of ${}^3 {\rm He}$ and
 the electron liquids.}
 
We believe that the physics of the strongly correlated electrons
in Si MOSFETs at $r_{s}\gg 1$ is quite similar to the low temperature
 physics of ${}^3{\rm He}$, despite the difference in interactions.
Indeed, in the absence of disorder, both systems exhibit a quantum
zero-temperature liquid-crystal 
transition. In both cases 
the transition is believed to be of the first order.
The liquid
phase near 
the transition in both systems
is characterized by a large ratio of the potential 
to kinetic energy. In the ${}^3{\rm He}$ liquid the
ratio between the Debye
energy $\Omega_{D}$ and the Fermi energy reaches 80 near the melting
pressure 
~\cite{nosier}. 
On the other hand, it is known, from numerical simulations \cite{cip},
that in the
two-dimensional
case the Fermi liquid-Wigner crystal transition takes place
at a relatively large
$r_{s}=r_{s}^{c}\sim 38$.
Therefore, in both systems there are wide regions of concentrations
where the systems are
in the liquid phases, though the potential energy is already much
larger than the kinetic one.

The only qualitative difference between 
these systems is that ${}^3{\rm He}$ crystallizes upon increasing 
the atomic concentration, while the electron system 
crystallizes when the concentration is lowered.

A model which describes properties of liquid ${}^3{\rm He}$ near
the crystallization
pressure 
has been introduced in~\cite{nosier,andreev,andreev1}.
It is based on the existence of the strong
inequality
$\hbar\Omega_{D}\gg E_{F(He)}$ which indicates that the
${}^3{\rm He}$ liquid can
be regarded as a system which is being close to
solid~\cite{nosier,andreev,andreev1}.
Since the system has strong short range crystalline order 
correlations, on a fast time scale, $\hbar/\Omega_{D}$ the atoms vibrate 
 near certain  
positions in their own cages. The system is locally rigid and can resist 
stress. On the much slower time 
scale $\tau_{(He)}\gg \Omega_{D}^{-1},\hbar/E_{F(He)}$, the cages
start
to
drift
around
each other leading to a plastic deformation expected in a liquid.
Here $E_{F(He)}$ is the Fermi energy of liquid ${}^3 {\rm
He}$.
The characteristic time $\tau_{(He)}$ is associated
with a collective rearrangement of "equilibrium"
particle configurations due to many-particle quantum tunneling 
~\cite{andreev,andreev1}. It 
determines all 
low temperature Fermi liquid parameters of the systems. 
At $t>\tau_{(He)}$ the quantum-mechanical indistinguishability of  
particles plays a decisive role in determining the ground state of the 
system. 
 The energy uncertainty $\hbar/\tau_{(He)}=
E_{d(He)}\ll E_{F(He)}$ in this case 
determines 
the order of magnitude of the temperature of the quantum degeneracy
$E_{d(He)}$. The quasiparticle
effective mass $m^{*}_{(He)}$ can be introduced in the usual way
$E_{d(He)}=p_{F(He)}^{2}/2m^{*}_{(He)}$. It can be extracted from the
temperature
dependence of the electron heat capacity. 
Accordingly, at $T=0$, both the effective mass 
$m^{*}_{(He)}\sim 
\hbar\tau_{(He)} n\gg m_{(He)}$, and the magnetic susceptibility of the
liquid 
$\chi_{L(He)}\sim n_{(He)}\mu^{2}_{N} \tau_{(He)}/\hbar\gg \chi_{0(He)}$
are
significantly enhanced as
compared
to their values $\chi_{0(He)}$ and $m_{(He)}$ in a noninteracting
liquid
\cite{andreev,andreev1}. Here $\mu_{N}$ and $n_{(He)}$ are the nuclear
Bohr magneton and the concentration respectively. The
latter formula corresponds to the Curie
 susceptibility with $T=\hbar/\tau_{(He)}$. It reflects the fact the
particles change their 
equilibrium positions on the time scale $\tau_{(He)}$.

According to the picture of the  liquid which
is nearly solid\cite{nosier}, not only the linear spin
susceptibility, but also all nonlinear ones should be large. It means that
the complete polarization of the electron liquid can be achieved at
relatively
low magnetic fields $\mu_{N} H^{c}\sim E_{d(He)}<E_{F(He)}$.

Furthermore, near the crystalization point the magnetic susceptibility
of the liquid ${}^3{\rm He}$ is
much smaller than the susceptibility of the solid ${}^3{\rm
He}$. Therefore
the
energy density
of the solid decreases with
$H$ faster that the energy density of the liquid.
This means that the magnetic field drives the
system toward the crystallization \cite{nosier,lal}.
 By itself it does not mean that the 
effective mass of
quasiparticles $m_{(He)}^{*}(H)$ increases with $H$.
However the calculations in the framework of the Hubbard model 
showed \cite{voll} that it is the case and
$m^{*}_{(He)}(H)-m^{*}_{(He)}(0)>0$, which
means that $\tau_{(He)}(H)$ is an increasing function of $H$.

Let us now discuss the $T$-dependence of the viscosity of the ${}^3{\rm
He}$
liquid $\eta_{(He)}(T)$.
At low temperatures $T\ll E_{d(He)}$ the viscosity $\eta_{(He)}(T)$ of
liquid 
${}^3 {\rm He}$
is given by the Fermi liquid theory $\eta_{(He)}\sim
T^{-2}$ ~\cite{lankin}.
There is, however, a wide temperature interval $E_{d}<T<\Omega_{D}$ where
the liquid is
nondegenerate already, but still it is strongly correlated.
Surprisingly enough the experimental data for the $T$ dependence of
$\eta_{(He)}$ in this temperature interval are unavailable. Theoretically
this dependence was
considered in \cite{andreev,andreev1} where it was argued that
in this temperature interval $\eta\sim T^{-1}$.

Let us turn now to the case of the electron liquid. If $r_{s}\gg 1$ we
have
$\Omega_{p}\gg E_{F}$, where $\Omega_{p}=(4\pi e^{2}n^{3/2}/m)^{1/2}$
is the plasma frequency at the wavelength of order of the interelectron
distance. Thus the discussed above model of strongly
correlated 
liquid
 can be applied to the
electron liquid as well.
Namely, one can view the electron liquid as a system which is close to
the Wigner crystallization. Due to
existence of the short range crystalline order effective electron mass
$m^{*}=\hbar n\tau$ and
the spin susceptibility $\chi=\mu^{2}\tau n/\hbar$ are enhanced compared
to
the noninteracting Fermi gas
case, where $\mu$ is the Bohr magneton. The enhancement of both
quantities is governed by the parameter $\tau\gg \Omega_{p}^{-1}$. 
Furthermore, since zero temperature spin susceptibility of the Wigner
crystal $\chi_{S}$ is
larger than the
susceptibility of the Fermi liquid $\chi_{L}$, the
energy density
of the Wigner crystal $E_{S}(H_{\|})=E_{S}(0)-\chi_{S}H^{2}_{\|}$
decreases with
$H_{\|}$ faster that the energy density of the Fermi liquid
$E_{L}(H_{\|})=E_{L}(0)-\chi_{L}H^{2}_{\|}$.
This means that the magnetic field parallel to the film drives the
electron
system toward the crystallization \cite{nosier,lal}.
 An estimate for the change of the critical electron concentration is
\begin{equation}
n_{c}(H_{\|})-n_{c}(0)=\frac{(\chi_{S}-\chi_{L})H_{\|}^{2}}
{\nu_{L}-\nu_{S}}>0
 \end{equation}
Here $\nu_{L}$ and $\nu_{S}$ are chemical potentials of the liquid and
crystal respectively.
By analogy with the theory of ${}^3{\rm He}$ we can assume that near the
crystalization point the effective quasiparticle mass in the
liquid $m^{*}(H)$ and the parameter $\tau(H)$ are increasing functions of
$H$. 

Concluding this section we would like to remind that the existence
of two different energy scales is not a unique property of the quantum
liquids near the solidification point. It has been known since Frenkel
\cite{frenk} that melting of ordinary classical liquids
occurs at temperatures much smaller
than the interaction energy between atoms. As a result,
following to \cite{frenk}, the viscosity of classical liquids
is governed by thermal-activation of the over-the-barriers processes
and decreases with $T$ exponentially.

To describe the temperature and the magnetic field dependence of the
resistance of the system we have to consider the electron system in the
presence of random elastically scattering potential.
Below we consider two limiting cases: a model of a
potential slowly waryng in space and a case when the potential is modeled
by short-range randomly
distributed scatterers.

 \section{The case of a smooth scattering potential.}

Let us consider a model where the fluctuations of the external potential
of a relatively small amplitude
 are smooth functions
of coordinates. If the electron concentration $n$ is close to the 
critical $n_{c}$
the system can get split into the regions of a Fermi
liquid and a Wigner
crystal.
This model can  explain the following experimental facts.

A. The
fractions of volume occupied by the Fermi liquid and the Wigner
crystal depend on $n$ and therefore the system should exhibit a 
percolation type zero temperature metal-insulator transition as 
 $n$ decreases and the area occupied by the Wigner crystal grows.

B. Since $\chi_{S}\gg \chi_{L}$,
 the magnetic field parallel to the film drives the
electron
system toward the crystallization \cite{nosier,lal} and  the
fraction of volume occupied by the Wigner
crystal increases with increasing $H_{\|}$. This
 leads to a big positive magnetoresistance as a function of $H_{\|}$.
The magnetoresistance should saturate when $H_{\|}>H_{\|}^{c}$ and the
electron Fermi liquid is polarized.
The saturation field $H^{c}_{\|}\sim \frac{E_{d}}{\mu}<E_{F}$
 decreases with decreasing $n$. This conclusion is in agreement with the
experiment~\cite{klap2}.

Depending on the value of
$n$ and the amplitude of the scattering potential
 at $H_{\|}>H_{\|}^{c}$ the system could be either in the metallic or in
the dielectric regime.
If $n<n^{c}(H_{\|}^{c})$ the system at $H_{\|}>H^{c}_{\|}$ is in the
insulating
regime. It exhibits a giant positive magnetoresistance which corresponds
to decreasing of the localization radius with increasing $H_{\|}$.

C. Consider the case $n<n_{c}(H^{c}_{\|})$ when at $H_{\|}>H_{\|}^{c}$ the
system is in the insulating phase. The
model presented above can also explain why in the presence of
the parallel magnetic field $H_{\|}>H_{\|}^{c}$ the system exhibits a big
negative magnetoresistance as a function of the component of the magnetic
field $H_{\bot}$ perpendicular to the film.
Indeed at $H_{\|}>H_{\|}^{c}$ electron spins are polarized and the
problem of the
magnetoresistance is of the single particle nature. 
 It was shown that at relatively small magnetic field when
$L_{H_{\bot}}\gg \xi$ it
is
 dominated by an interference of direct tunneling paths
\cite{shkl,shkl1,feng}. Here
$L_{H_{\bot}}=\sqrt{c\hbar/e H_{\bot}}$ is the magnetic length and
$\xi$ is the localization
radius. In this case the magnetoresistance is shown to be negative and 
the $H_{\bot}$ dependence of the resistance corresponds to a
correction to the localization radius $(\xi(H_{\bot})-\xi(0))\sim
\xi^{2}(0)L_{H_{\bot}}^{-1}$.
Existence of such an interference effect depends
on the  spin structure of the system ground state \cite{shkl1,spiv1}.
At $H_{\|}=0$ the tunneling
is a collective process which
 involves an interchange of
positions of electrons with different spins. As a result, final
 states of the system, which
correspond to different tunneling paths have different spin
configurations and, therefore, they are orthogonal. As a result, the
single particle interference mechanism of the negative magnetoresistance
can be significantly suppressed in the case $H_{\|}=0$.

D. The significant increase of the resistance as a function of temperature
can also be explained naturally in the framework of the presented above
model as a consequence of the Pomeranchuk effect. Indeed, if temperature
is
not significantly smaller than the spin
exchange energy in the Wigner crystal, the spin entropy of the Wigner
crystal
is larger than the entropy of the Fermi liquid. This means that the
Wigner crystal regions grow with increasing temperature.

E. The Pomeranchuk effect disappears when $H_{\|}>H^{c}_{\|}$ and electron
spins
are polarized. In this case entropies of both  the liquid and the solid
are much smaller than the spin entropy of the crystal at $H_{\|}=0$.
This means that in the first approximation the areas occupied by the
crystal and the liquid are $T$-independent. 
This explains the fact that in the
metallic
state at $H_{\|}>H_{\|}^{c}$ the $T$-dependence of the resistance
 is much smaller than in the case
$H_{\|}=0$ \cite{vit,klap1}.

F. The question about the origin of the anisotropy of the
magnetoresistance in the metallic phase is open. The fact that 
the magnetoresistance in the parallel field is larger than the
magnetoresistance in the perpendicular field 
implies that the orbital
effects are insignificant and the magnetoresistance is determined by spin
magnetization of the system.
 A possible explanation of the large anisotropy of the spin magnetization
induced by the magnetic field
can be
related to the
existence of the Rashba term in one-particle spectrum of electrons
\begin{equation}
\alpha\bbox{p}[\bbox{\sigma}\times \bbox{n}]
\end{equation}
 Here $\bbox{n}$ is a unit vector
normal to the conducting plane and $\bbox{\sigma}$ is the vector of Pauli
matrices and $\bbox{p}$ is the electron momentum.
A reliable estimate for the value of $\alpha$ in Si-MOSFETs is
unavailable at the moment. A rough estimate ~\cite{pudalovrash} implies
that
the
magnitude of
Eq.2 can be just several times less than the electron Fermi energy.
We would like to note, however, that a) in the presence of the
electron-electron interaction the degeneracy energy is renormalized
($E_{d}<E_{F}$) and b)the electron-electron interaction renormalizes the
factor $\alpha$ in Eq.2~\cite{raikh}. It is natural to expect that the
latter renormalization
is determined by the parameter $\tau\Omega_{p}\gg 1$. Both effects lead to
increasing of the relative amplitude of the Rashba term. At last, we would
like to mention that the spin-orbit interaction can manifest itself also
in the in-plane anisotropy when the resistance depends on the relative
orientation of the current and the magnetic field parallel to the plane
~\cite{pudanis,raikh1}.

At last we would like to mention that the resistance of order
$\hbar/e^{2}$ is a typical feature of the percolation transition. (See for
example ~\cite{meir}.) The difference between the presented above model
and 
~\cite{meir} is in 
the many particle nature of the transition which, for example, makes it
sensitive to the parallel magnetic field. 

A difficulty
associated with the model of smooth potential is the following. In the
case of a small amplitude
of the fluctuations of the scattering potential 
at $T=0$ the transition should take place 
at $r_{s}=r_{s}^{c}\sim 38$, which is significantly larger than the
experimental
value $r^{c}_{s}=10-20$.
On the other hand, if the amplitude of the potential is large, in
principle, a percolation type transition can take place at any $r_{s}$.
In the latter case, however, the system breaks into three phases:
insulator, Wigner crystal and the Fermi liquid. 
One  can neglect the difference in
compressibilities of the Fermi
liquid and the Wigner crystal. As a result, the electrostatic analysis
shows that 
the electron density should
 increase as a square root of the
distance from the boundary~\cite{shklovskii}. Then the assumption that the
liquid-crystal transition takes place at $r_{s}=38$ leads to a conclusion
that the fraction of the area occupied by the Wigner crystal is
numerically small and, therefore, at $r_{s}\sim 10-20$ the discussed above
$T$ and
$H$ dependences of the resistance should not be pronounced.
 A possibility to have $r^{c}_{s}$ less than $38$
in the presence of short range scattering potential is considered in the
next section.

We would also like to mention that even in
the framework of the presented above model the question whether there is a
well defined transition or just a sharp crossover between metallic and
insulating phases is oppen.
Indeed, in the two-dimensional case an arbitrarily
small disorder destroys the  first
order phase transition
 \cite{imre}.
The question of whether in this case the first order Fermi
liquid-Wigner
crystal transition is transformed into a second order one, or it is
destroyed
completely is open. (See a discussion of this question in \cite
{kiv}.)

The question about
possible localization in the ``metallic phase" in the presence of a weak
disorder is also open.
All existing theories of the localization
are perturbative in electron-electron interaction.
At $r_{s}\gg 1$ the potential energy is much
larger than the kinetic energy and these theories are not reliable
especially in the case $r_{s}\gg 1$ when the renormalized mass
is big, which means
that the size of electron packets which carry current should be large
as well. 
 It may be that
the localization length in the metallic phase is larger than the sample
size.

\section{Short range scattering potential.}

In this section we show that a) in the presence of a short range potential
the
Wigner crystallization takes place at $r_{s}<38$ and b) one can
explain the
mentioned above experiments using the fact that that the electron
liquid is nearly solid, but without involving the Wigner crystal state.
 
Suppose that there is an impurity with a short-range potential of a radius
$a\sim
n^{-1/2}$, which is embedded
into the metal with $r_{s}\gg 1$.
Let us now discuss the $r_{s}$ and $H_{\|}$ dependences of the
two-dimensional 
cross-section
of the quasiparticle elastic scattering on impurities $A(T=0)$ at
zero temperature.
 It has been pointed out~\cite{spivak} that due to the existence of 
the short range crystalline order in the Fermi liquid with $r_{s}\gg 1$ 
a large characteristic length $\tau \Omega_{p} n^{-1/2}\gg n^{-1/2}$
should
exists, where the liquid behaves like a solid.
Since the impurity pins the liquid, the value of $A$ should increase with
$\tau$ as well
\begin{equation}
A(T=0)\sim a\frac{m^{*}}{m}\gg a.
\end{equation}
Therefore, it is plausible that the "transition" takes place when
the impurity concentration $N_{i}$ equals to a critical one
\begin{equation}
N_{ic}=\frac{1}{A^{2}}\sim \frac{1}{a^{2}}(\frac{m}{m^{*}})^{2}
\end{equation}
 and strongly correlated regions near impurities overlap.
 Thus in disordered
samples
the ``transition" takes
place at $r_{s}\ll 38$ and the critical electron concentration decreases
with
increasing of the amount of disorder. This is in agreement with the
experimental observation \cite{shah,pudcam} that the critical
concentration
$n_{c}$
is lower in samples with higher mobility.
 We would like to note that numerical
simulations in the presence of disorder give a critical value
of
 $r_{s}$ for the Fermi liquid-Wigner crystal transition, which is
significantly smaller than 38~\cite{cip1,jlp}.

As it has been mentioned above both $\tau(H_{\|})$ and $m^{*}$ are
increasing functions of $H_{\|}$. As a result, the zero temperature 
resistance of the system
\begin{equation}
\rho(T=0)=\frac{\hbar}{e^{2}}
\frac{N_{i}A(T=0)}{n^{1/2}}.
\end{equation}
increases with $H_{\|}$ and saturates at $H_{\|}>H^{c}_{\|}$.

Temperature corrections to Eq.~1 should be small as long as $T\ll T_{d}$.
Since processes of electron-electron scattering in semiconductors
conserve the
total momentum, the $T$-dependence of the resistivity of the system
can be only due to
$T$-dependence of $A(T)$.
 
It has been shown~\cite{dolgopolov,dolgopolov1,sarma} that at low
temperatures the resistance of the
system increases linearly with $T$. This effect is due to the temperature
dependence of the Friedel oscillations induced by impurities.
If $r_{s}>1$ the value of the scattering cross-section turns out to be an
increasing function of
$H_{\|}$ as well~\cite{dolgopolov2}. This mechanism, in principle, could
explain
qualitatively $T$ and $H_{\|}$ dependences of the resistance of the
metallic state. It is hard, however, to explain in
the framework of~\cite{dolgopolov,dolgopolov1,dolgopolov2} why at
$H_{\|}>H_{\|}^{c}$ the temperature
dependence of the resistance in the metallic state is much smaller than at
$H=0$ and why the
magnetoresistance in the dielectric state with respect to $H_{\bot}$ is
negative.  
It may be, though, that the mechanism of the $T$ dependence considered
in~\cite{dolgopolov,dolgopolov1,dolgopolov2,sarma} is relevant in
the case of
$GaAs$ samples.

Let us now discuss the $T$-dependence of the resistance in the temperature
interval $T_{d}<T<\Omega_{p}$.
 Though in this
case the
liquid is not degenerate, it is strongly correlated. Therefore the
electron-electron
scattering in
the liquid is very effective and the local
equilibrium is reached in a short time on a spatial scale of order
$n^{-1/2}$.
 As a result, the flow
of the liquid near an impurity can be considered in the framework of
hydrodynamics. In the two-dimensional case the electron liquid exerts a
force on an
impurity given by the Stokes
formula $F\sim \eta u/\ln(\eta/nua)$, where $u$ and $\eta$ are the liquid
hydrodynamic velocity and viscosity of the electron liquid 
respectively. In a system with a finite concentration of impurities
the logarithmic factor in 
the equation for $F$ should be substituted for $\ln(1/aN_{i}^{1/2})$. 
Thus the resistance of the system is 
\begin{equation}
\rho(T)\sim \frac{N_{i}\eta(T)}{e^{2}n^{2}}
\ln^{-1}\frac{1}{N_{i}^{1/2}a},
\end{equation}
 Strictly speaking the hydrodinamic
approach is valid when $a>l_{ee}$. Here $l_{ee}$ is the electron-electron
mean free path. However, since $a$ dependence in
Eq.6 is only logarithmic, the estimate Eq.6 is also valid in the case of
semi-quantum liquid when $a\sim
l_{ee}\sim n^{-1/2}$. The logarithmic factors in Eq.6 is
associated with the Stokes paradox in the two-dimensional case
\cite{landau}.
It is not universal. For example, electron-phonon scattering will change
it significantly. We will neglect this factor estimating the resistance of
the system by the order of magnitude.

 The essential
ingredient of the picture which leads to the Stokes formula
for $F$ and to Eq.6 is that the hydrodynamic
velocity near impurities is significantly reduced as compared to its 
bulk
value. This is the reason why the conventional description of the
two-dimensional electron
system with the help of the Boltzmann kinetic equation containing the
electron-electron and the electron-impurity scattering integrals
\cite{levinson} would give 
a result completely different from Eq.6. In the former case the electron
distribution function is spatially uniform and, therefore, the resistance 
of the system is independent of $l_{ee}$ and proportional to the
electron-impurity scattering rate.

The viscosity of the electron liquid in the
semi-classical regime can be estimated in a way similar to
\cite{andreev,andreev1}.
On a time scale smaller than
$\tau$ the structure of the low energy excitations in the liquid is
similar to the structure of excitations in glasses \cite{chandra}: At $T\ll
\hbar\Omega$ the liquid excitations are two-level systems with the
density of states per particle $\nu_{0}\sim U^{-1}$. In the
case 
of the electron liquid $U\sim e^{2}\epsilon^{-1}n^{1/2}$ is the typical
interelectron 
interaction energy.
As a result \cite{andreev}, contrary to the gas case, the viscosity of
the liquid 
$\eta\sim \hbar n\nu_{0} V^{2}/T$
 decreases with increasing
temperature. Here $V$ is the typical matrix element of the transition
between 
the states in the two level systems. 
Making a natural assumption that
$V\sim \hbar\Omega_{p}$ we get 
\begin{equation}
\eta(T)\sim \frac{T_{d}}{T}\frac{m^{*}}{m}
\hbar n.
\end{equation}
Thus, we arrive to the conclusion that at $T>T_{d}$ the resistance
associated with the impurity scattering
\begin{equation}
\rho(T)\sim
\frac{\hbar}{e^{2}}\frac{N_{i}}{n}\frac{m^{*}}{m}\frac{T_{d}}{T}
\end{equation}
should decrease with increasing $T$. Therefore, $\rho(T)$ may have a
maximum at $T\sim
T_{d}$.
At $T=0$ and in the "metal" regime   
$(N_{i}A^{2}(T=0)\ll 1)$ contributions from different
impurities into $\rho(T)$ are independent and
\begin{equation}
\rho(T=0)=\frac{\hbar}{e^{2}}
\frac{N_{i}A(T=0)}{n^{1/2}}.
\end{equation}
 Thus at $T=T_{d}$ Eq.~4
matches
the zero
temperature value of
the impurity scattering cross-section Eq.~3.

\section{conclusion}

We would like to mention several consequences of the model presented above 
 which can be checked experimentally.

1. In the metallic phase at $r_{s}\gg 1$ the thermopower,
normalized by $\rho(T=0)$, should be relatively large due to the 
strong enhancement of the effective mass $m^{*}/m\gg 1$. 
Also, the thermopower should be a strongly increasing function
of $H_{\|}$ because of the corresponding increase in $m^{*}(H_{\|})$. 
In principle, large thermopower in a metal can also occur
due to localized spins present in the sample and the Kondo effect 
associated with them.
However, in the latter case the
thermopower would be a decreasing function of $H_{\|}$.

2. The tunneling density of states in
the 
metallic regime with $r_{s}\gg 1$ is reduced by the factor 
$m^{*}(H_{\|})/m\gg 1$ compared to that of noninteracting electrons. 
The tunneling density of states
will be additionally suppressed
by the magnetic field parallel to the film. 

3. In the Wigner crystal state, as well as in the liquid with $r_{s}\gg
1$, the electron compressibility is negative, whereas at $r_{s}\ll 1$ it
is positive~\cite{belo}. The magnetic field applied parallel to the 
plane drives the system toward the Wigner crystal state,
and thus it should make the compressibility of the ``metallic state" 
at $38\gg r_{s}\gg 1$ even more negative.

4.Following the presented above picture the saturation of the
magnetoresistance in the metallic phase in the parallel magnetic field 
takes place at
$H_{\|}>H_{\|}^{c}$ when the electron Fermi liquid gets polarized.
On the other hand the Pomeranchuk effect disappears when the Wigner
crystal is polarized. This means that the significant temperature
dependence of the metallic state may disappear in a much smaller magnetic
field
$H_{\|}\sim \frac{T}{\mu}\ll H^{c}_{\|}$.

This work was supported by Division of Material Sciences, 
U.S.National Science Foundation under Contract No. DMR-9205144.
We would like to thank A.F.~Andreev, S.~Chakravarty,
A.~Finkelshtein, D.~Feldman, T.~Gimarchi, D.~Khmelnitskii, K.~Kikoin,
S.~Kivelson,S.~Kravchenko,
A.I.~Larkin, L.~Levitov, V.~Pudalov, J.L.~Pichard, M.~Raikh, M.~Sarachik,
B.~Shklovskii, S. Das Sarma and F.~Zhou
for useful discussions.~

\end{multicols}

\end{document}